\documentclass[pra,aps,amsmath,amssymb,showkeys,eqsecnum]{revtex4}
\usepackage{graphicx}
\newcommand{\hh}{{\mathcal{H}}}

\newcommand{\pen}{\openone}
\newcommand{\bro}{\boldsymbol{\rho}}

\newcommand{\me}{{\mathsf{E}}}

\newcommand{\cald}{{\mathcal{D}}}
\newcommand{\calp}{{\mathcal{P}}}
\newcommand{\cle}{{\mathcal{E}}}
\newcommand{\clf}{{\mathcal{F}}}

\newcommand{\xdif}{{\mathrm{d}}}

\newcommand{\rmh}{{\mathrm{H}}}
\newcommand{\psyms}{{\mathsf{\Pi}}_{\mathrm{sym}}^{(s)}}
\newcommand{\psymt}{{\mathsf{\Pi}}_{\mathrm{sym}}^{(t)}}

\newcommand{\tr}{\mathrm{tr}}

\newcommand{\wta}{{\tilde{a}}}
\newcommand{\wb}{{\tilde{b}}}
\newcommand{\vcp}{{\mathtt{p}}}
\newcommand{\veps}{\varepsilon}

\begin{document}
\clearpage
\preprint{}

\title{Estimating the Shannon entropy and (un)certainty relations for design-structured POVMs}

\author{Alexey E. Rastegin}
\affiliation{Department of Theoretical Physics, Irkutsk State University, 1 K. Marx St., Irkutsk 664003, Russia}

\begin{abstract}
Complementarity relations between various characterizations of a
probability distribution are at the core of information theory. In
particular, lower and upper bounds for the entropic function are of
great importance. In applied topics, we often deal with situations,
where the sums of certain powers of probabilities are known. The
main question is how to convert the imposed restrictions into
two-sided estimates on the Shannon entropy. It is addressed in two
different ways. The more intuitive of them is based on truncated
expansions of the Taylor type. Another method is based on the use of
coefficients of the shifted Chebyshev polynomials. We propose here a
family of polynomials for estimating the Shannon entropy from below.
As a result, estimates are more uniform in the sense that errors do
not become too large in particular points. The presented method is
used for deriving uncertainty and certainty relations for positive operator-valued measures
assigned to a quantum design. Quantum designs are currently the
subject of active researches due to potential use in quantum
information science. It is shown that the derived estimates are
applicable in quantum tomography and detecting steerability of
quantum states.
\end{abstract}


\maketitle

\pagenumbering{arabic}
\setcounter{page}{1}

\section{Introduction}\label{sec1}

The concept of entropy is fundamental in statistical physics and
information theory \cite{wehrl78}. Irrespective of the context of
their use, entropic functions are hardly exposed to measure
immediately. It is not obvious that the total uncertainty about a
multipartite system corresponds to the sum of the entropies of its
parts \cite{renner20}. Hence, we are interested in ways to connect
entropic quantities with those characteristics that are easier to
observe in practice. Results of such a kind constitute an essential
part of information sciences, including quantum area
\cite{fuchs96,wilde17}. Inequalities between the entropy and the
index of coincidence are one of the examples naturally raised in several
ways \cite{harr2001}. Connecting inequalities between several
information measures can also be used to choose properly a single
notion appropriate in a particular context. The authors of
\cite{fuchs99} reviewed this issue with respect to quantum
cryptography. In this paper, the problem of entropy characterization
will be treated within uncertainty and certainty relations for positive operator-valued measures (POVMs)
assigned to a quantum design.

Information entropies are useful to characterize genuine
uncertainties in various situations \cite{rudolph14}. They give a
natural way to pose uncertainty relations in the presence of quantum
memory \cite{bccrr10}. For discrete observables, the entropic
approach to uncertainty relations was developed in
\cite{deutsch,maass}. Basic results found in this area are reviewed
in \cite{ww10,cbtw17}. For entropic uncertainty relations with
continuous variables, see \cite{brud11,cerf19} and references
therein. In quantum information science, special types of
measurements have found a lot of attention. Mutually unbiased bases
\cite{bz10} and symmetric informationally complete measurements
\cite{rbksc04} are especially important examples. Quantum designs,
also called complex projective designs, have been considered for
several reasons \cite{scottjpa,ambain07}. Characterizing
uncertainties in design-structured POVMs is one of the questions raised
in this connection. The authors of \cite{guhne20} formulated
uncertainty relations for such POVMs in terms of the R\'{e}nyi and
Tsallis entropies. R\'{e}nyi formulation was improved in
\cite{rastdes}. The results of \cite{guhne20,rastdes} also give
entropic inequalities to detect a special kind of nonlocal
correlations known as quantum steerability.

As the R\'{e}nyi entropy cannot increase with growth of its order,
uncertainty relations derived in \cite{guhne20,rastdes} imply
estimates on the corresponding Shannon entropy from below. However,
such estimates are sufficiently far from optimality. The question of
deriving uncertainty relations in terms of the Shannon entropy is
beyond the methods of \cite{guhne20,rastdes}. The aim of this work
is to address this question in detail. Moreover, the developed
approach naturally leads to estimates on the corresponding Shannon
entropy from above. Thereby, certainty relations for
design-structured POVMs are formulated. The paper is organized as
follows. In Section \ref{sec2}, the methods to get two-sided
estimates on the Shannon entropy are considered. Here, we propose
polynomials whose coefficients are not determined according to
Taylor's scheme. Section \ref{sec3} is devoted to uncertainty and
certainty relations for POVMs assigned to a quantum design. These
relations are formulated and compared with previous results within
several examples. Applications for estimating the von Neumann
entropy and steering inequalities are briefly mentioned. Section
\ref{sec4} concludes the paper. Some auxiliary material is presented
in two appendices.

\section{On two-sided estimating on the Shannon entropy}\label{sec2}

In this section, we present two-sided estimates on the Shannon
entropy in terms of the power sums of probabilities. The Shannon
entropy of probability distribution $\vcp=\{p_{j}\}$ is defined as
\cite{cover2006,hay2017}
\begin{equation}
H_{1}(\vcp):=-\sum\nolimits_{j} p_{j}\ln{p}_{j}
\, . \label{shdef}
\end{equation}
For a pair of discrete random variables, the conditional entropy of
$X$ given $Z$ reads as \cite{cover2006,hay2017}
\begin{equation}
H_{1}(X|Z):=\sum\nolimits_{z}p(z)\,H_{1}(\vcp_{X|z})
\, , \label{csea}
\end{equation}
where $\vcp_{X|z}$ consists of conditional probabilities $p(x|z)$.
The authors of \cite{harr2001} introduced information diagrams that
allow them to study relations between (\ref{shdef}) and the index of
coincidence:
\begin{equation}
I^{(2)}(\vcp):=\sum\nolimits_{j} p_{j}^{2}
\, . \label{ic2def}
\end{equation}
It is natural to generalize (\ref{ic2def}) as
\begin{equation}
I^{(s)}(\vcp):=\sum\nolimits_{j} p_{j}^{s}
\, . \label{icsdef}
\end{equation}
Such indices were briefly discussed in \cite{harr2001}. We will use
them with positive integer $s$. For $s=0$, the index (\ref{icsdef})
is equal to the number of nonzero probabilities. The Tsallis
$s$-entropy is defined as \cite{tsallis}
\begin{equation}
H_{s}(\vcp):=\frac{1}{1-s}\left(\sum\nolimits_{j} p_{j}^{s}-1\right)
=\frac{I^{(s)}(\vcp)-1}{1-s}
\ . \label{tsadf}
\end{equation}
In the limit $s\to1$, the latter reduces to (\ref{shdef}). For a
discussion of basic properties of (\ref{tsadf}) and other entropic
functions, see section 2.7 of \cite{bengtsson}.

Suppose that several quantities of the form (\ref{icsdef}) are known
exactly. Hence, available values of the Shannon entropy have to be
restricted. The main question is how to characterize allowed
entropic values explicitly. So, one way to express two-sided
estimates on the entropic function is in terms of indices of the form
(\ref{icsdef}). In this work, we aim to build a polynomial $q(x)$ of
degree $n$ such that $q(0)=0$ and
\begin{equation}
x\ln{x}\leq{q}(x)=\sum\nolimits_{s=1}^{n} q_{n}^{(s)}x^{s}
 \label{fipoq}
\end{equation}
for all $x\in[0,1]$. The inequality (\ref{fipoq}) allows us to
estimate the Shannon entropy from below. Also, we integrate both the
sides of the inequality
\begin{equation}
1+\ln\tilde{x}\leq1+\sum\nolimits_{s=1}^{n} q_{n}^{(s)}\tilde{x}^{s-1}
 \label{fipoq1}
\end{equation}
from $\tilde{x}=x$ to $\tilde{x}=1$, whence
\begin{equation}
-x\ln{x}\leq1+\sum_{s=1}^{n}\frac{q_{n}^{(s)}}{s}-x-\sum_{s=1}^{n}\frac{q_{n}^{(s)}x^{s}}{s}
\ . \label{fipoq2}
\end{equation}
This inequality will be used to estimate the Shannon entropy
from above. In effect, both the relations (\ref{fipoq}) and
(\ref{fipoq2}) should be applied with rescaled probabilities.

The main question is to find desired polynomials. At first glance,
spline functions may seem an appropriate tool in this situation.
However, this approach is hardly applicable to the case of interest.
Using suitable truncation of Taylor series is a natural way with
rich history. Deficiencies of this approach come from the fact that
truncated series gave very accurate results near the point of
expansion but insufficient ones in peripheral regions
\cite{lanczos}. Nevertheless, truncated expansions of the Taylor
type are still a powerful tool for studying inequalities of
interest. Instead, there are power expansions whose coefficients are
not determined according to the Taylor scheme. In numerical
analysis, such expansions are often based on the use of a suitable
family of orthogonal polynomials. The significance of Chebyshev
polynomials in various problems of applied analysis was clearly
demonstrated by Lanczos \cite{lanczos}. Before addressing his ideas,
we consider truncated expansions of the Taylor type.

Putting $1-x=z\in[0,1]$ and using well-known expansions, we have
\begin{equation}
\ln{x}=\ln(1-z)=-\sum_{r=1}^{\infty}\frac{z^{r}}{r}
\leq-\sum_{r=1}^{n-1}\frac{z^{r}}{r}
\, . \label{ntayn}
\end{equation}
Multiplying the latter by $-x\leq0$, one gets
\begin{equation}
f_{n}(x)\leq-x\ln{x}
\, , \label{tayln0}
\end{equation}
where
\begin{equation}
f_{n}(x)=x\sum_{r=1}^{n-1}\frac{(1-x)^{r}}{r}
=\sum_{s=1}^{n} a_{n}^{(s)}x^{s}
\, . \label{fndef}
\end{equation}
It is immediate to check that
\begin{equation}
a_{n}^{(1)}=\sum_{r=1}^{n-1}\frac{1}{r}
\ , \qquad
a_{n}^{(s)}=(-1)^{s-1}\sum_{r=s-1}^{n-1}\frac{1}{r}\>\binom{r}{s-1}
\qquad
(2\leq{s}\leq{n})
\, . \label{aeff}
\end{equation}
To estimate $(-x\ln{x})$ from above, we could combine
$x\ln{x}\leq-f_{n}(x)$ with (\ref{fipoq}) and (\ref{fipoq2}). The more
direct way is to truncate the expansion of $(z-1)\ln(1-z)$. In any
case, the result reads as
\begin{equation}
-x\ln{x}\leq{h}_{n}(x)
\, , \label{tayln1}
\end{equation}
where
\begin{equation}
h_{n}(x)=(1-x)\biggl(1-\sum_{r=1}^{n-1}\frac{(1-x)^{r}}{r(r+1)}\biggr)
=\sum_{s=0}^{n} b_{n}^{(s)}x^{s}
\, . \label{hndef}
\end{equation}
The coefficients are expressed as $b_{n}^{(0)}=\frac{1}{n}$,
\begin{equation}
b_{n}^{(1)}=\sum_{r=2}^{n-1}\frac{1}{r}
\, , \qquad
b_{n}^{(s)}=\frac{(-1)^{s-1}}{s}\sum_{r=s-1}^{n-1}\frac{1}{r}\>\binom{r}{s-1}
\qquad
(2\leq{s}\leq{n})
\, . \label{boeff}
\end{equation}
As obtained due to the Taylor scheme, the estimates (\ref{tayln0})
and (\ref{tayln1}) demonstrate a typical behavior. They give a very
good approximation from their own side in some left neighborhood of the
point $x=1$. In contrast, inaccuracies become comparatively large in
a right neighborhood of the point $x=0$. They are most obvious for
(\ref{tayln1}) due to $h_{n}(0)=1/n$. Thus, there is a certain
asymmetry with respect to estimating in left and right regions of
the range $x\in[0,1]$. This asymmetry can be reduced due to the
methods described in \cite{lanczos}. Although the inequalities
(\ref{tayln0}) and (\ref{tayln1}) hold for $n=1$, it is not used in
the following.

Using orthogonal polynomials for approximation, expansions with
rigid coefficients are best known. That is, adding a next member of
the chosen orthogonal family into our expansion does not affect the
sum we have obtained before. As a reverse side of this fact, we have
to deal with oscillating approximations. One can only minimize an
integral error, say, in the sense of least squares. Such expansions
are not suitable to fit the function of interest from above or below
solely. As was emphasized by Lanczos (see, e.g., chapter VII of
\cite{lanczos}), the condition of rigidity of coefficients received
an exceptional attention for rather historical reasons. To reach
more effective approximations, adding a next member should be
combined with changing all the coefficients. So, we have arrived at
expansions with flexible coefficients.

In section VII.12 of his book \cite{lanczos} Lanczos described the
so-called $\tau$-method to solve differential equations. The next
paragraph of \cite{lanczos} is devoted to a reformulation with
canonical polynomials. Let function $x\mapsto{y}(x)$ obey a certain
ordinary differential equation. Its power fitting can be found as an
approximate solution of this equation. When the original equation is
not solvable in polynomials, it is modified by adding an
inhomogeneous term. Lanczos proposed to add a term proportional to
some shifted Chebyshev polynomial. This procedure generally includes
a certain freedom. The following example gives a useful power
expansion of $y(x)=x\ln{x}$. For the equation
\begin{equation}
xy^{\prime}(x)-y(x)=x
 \label{oden}
\end{equation}
with the boundary condition $y(1)=0$, one adds $n$th shifted
Chebyshev polynomial $T_{n}^{*}(x)$ on the right. Some facts about
these polynomials are recalled in Appendix \ref{cheb}. The procedure
results in the sum
\begin{equation}
\frac{(-1)^{n}}{2n^{2}}\biggl(x-1+\sum_{s=2}^{n} c_{n}^{(s)}\,\frac{x^{s}-x}{s-1}\biggr)
\, , \label{lansun}
\end{equation}
which gives a polynomial approximation of $y(x)=x\ln{x}$
\cite{lanczos}. Since it oscillates around the function to be
fitted, we shall modify (\ref{lansun}). Indeed, the latter does not
vanish at the point $x=0$. For odd $n$, it takes the value
$1/(2n^{2})>0$ for $x=0$, though the initial function is negative
for all $x\in(0,1)$. The constant term in (\ref{lansun}) is
determined by $T_{n}^{*}(x)$ on the right, whereas any term
$\propto{x}$ is a solution of homogeneous equation. Adding to
(\ref{lansun}) the linear term $\frac{(-1)^{n}(1-x)}{(2n^{2})}$, for
$n\geq2$ we finally write
\begin{equation}
g_{n}(x)=\frac{(-1)^{n}}{2n^{2}}\,\sum_{s=2}^{n} c_{n}^{(s)}\,\frac{x^{s}-x}{s-1}
 \label{gansun}
\end{equation}
so that $g_{n}(0)=0$ and $g_{n}(1)=0$. Polynomials of the form
(\ref{gansun}) are the first main finding of this work. They will be
used to estimate $x\ln{x}$ from above.

Let us consider the first derivative at the least points of the
interval $x\in[0,1]$. It follows from (\ref{gansun1}) that
$g_{n}^{\prime}(1)=1$ for even $n\geq2$ and
$g_{n}^{\prime}(1)=1-1/n^{2}$ for odd $n\geq3$. To express
$g_{n}^{\prime}(0)$ as an expression with definite sign, we use
(\ref{gansun0}) and (\ref{twoser}), whence
\begin{align}
g_{n}^{\prime}(0)&=-\frac{4}{n^{2}}\sum_{r=1}^{\lfloor{n}/2\rfloor} \frac{4r^{2}}{n-2r}
&(n\ \mathrm{odd})
\, , \label{gp0odd}\\
g_{n}^{\prime}(0)&=-\frac{4}{n^{2}}\sum_{r=1}^{\lfloor{n}/2\rfloor} \frac{(2r-1)^{2}}{n-2r+1}
&(n\ \mathrm{even})
\, . \label{gp0even}
\end{align}
To fit the original function, the line of $g_{n}(x)$ should go
closely to the line of $y(x)$. As $y^{\prime}(x)=1+\ln{x}\to-\infty$
for $x\to+0$, the graph of $y(x)$ tends to enter the abscissa axis
normally. Thus, the value $g_{n}^{\prime}(0)$ is expected to be
negative. On the other hand, this value is certainly finite, whence
the slope of tangential line is far enough from the vertical. In
some right vicinity of the point $x=0$, our approximation has to be
relatively poor, since derivatives are inevitably finite for any
polynomial. With growth of $n$, the absolute value of
$g_{n}^{\prime}(0)$ increases with improving an approximation. We
see from $-\infty<g_{n}^{\prime}(0)<0$ and $g_{n}(0)=y(0)$ that
$g_{n}(x)$ exceeds $y(x)$ in a right neighborhood of the point
$x=0$. The main result of this section is posed as follows.

\newtheorem{shn01}{Proposition}
\begin{shn01}\label{res01}
Let polynomials $g_{n}(x)$ be defined by (\ref{gansun}). For $x\in[0,1]$ and $n\geq2$, it holds that
\begin{equation}
x\ln{x}\leq{g}_{n}(x)
\, . \label{yleqg}
\end{equation}
\end{shn01}

{\bf Proof.}
We aim to prove nonnegativity of the difference
$\gamma_{n}(x)=g_{n}(x)-x\ln{x}$. It is zero for $x=0$ and $x=1$.
The task is to show (\ref{yleqg}) for points between $0$ and $1$.
First, the function $y(x)=x\ln{x}$ obeys (\ref{oden}). Combining
(\ref{gansun}) with (\ref{cn01}) immediately leads to
\begin{equation}
xg_{n}^{\prime}(x)-g_{n}(x)=\frac{(-1)^{n}}{2n^{2}}\,\bigl[\,T_{n}^{*}(x)-(-1)^{n}\bigr]+x
\, . \label{goden}
\end{equation}
Due to (\ref{oden}) and (\ref{goden}), one obtains
\begin{equation}
\frac{\xdif}{\xdif{x}}\>\frac{\gamma_{n}(x)}{x}=
\frac{x\gamma_{n}^{\prime}(x)-\gamma_{n}(x)}{x^{2}}=\frac{(-1)^{n}\,T_{n}^{*}(x)-1}{2n^{2}x^{2}}
\ . \label{mgoden}
\end{equation}
Putting $x=\cos^{2}\theta/2$ with $\theta$ between $0$ and $\pi$, we
have $T_{n}^{*}(x)=T_{n}(\cos\theta)=\cos{n}\theta$. So, the
right-hand side of (\ref{mgoden}) cannot be positive. For
$x\in(\veps,1)$ with small $\veps>0$, the ratio $\gamma_{n}(x)/x$ is
a smooth decreasing function vanishing at the right least point of
the interval. By monotonicity, this ratio is nonnegative for all
$x\in(\veps,1)$. Making $\veps>0$ arbitrarily small completes the
proof.
$\blacksquare$

Substituting $n=2$ gives $g_{2}(x)=-f_{2}(x)=x^{2}-x$ so that
$x\ln{x}\leq{g}_{2}(x)$ follows from (\ref{tayln0}). For other
values of $n$, functions of the form (\ref{gansun}) do not relate to
the Taylor scheme. The coefficients $c_{n}^{(s)}$ increase
considerably with growth of $n$. It is generally suitable to
consider only moderate values of $n$. With $n=15$, the difference
between the sides of (\ref{yleqg}) is less than one thousandth of
the maximum of $|x\ln{x}|$ in the interval $x\in[0,1]$. At fixed
$n$, the polynomial $g_{n}(x)$ improves an approximation for
relatively small $x$ without introducing considerable errors in a
left vicinity of the point $x=1$.

Due to (\ref{yleqg}), we estimate the function of interest from
below as
\begin{equation}
\sum_{s=1}^{n}\wta_{n}^{(s)}x^{s}\leq-x\ln{x}
\, . \label{wtaes}
\end{equation}
Here, the coefficients are expressed in terms of coefficients of
$n$th Chebyshev polynomial by the formulas
\begin{equation}
\wta_{n}^{(1)}=
\frac{(-1)^{n}}{2n^{2}}\,\sum_{s=2}^{n} \frac{c_{n}^{(s)}}{s-1}
\ , \qquad
\wta_{n}^{(s)}=\frac{(-1)^{n+1}}{2n^{2}}\,\frac{c_{n}^{(s)}}{s-1}
\qquad
(2\leq{s}\leq{n})
\, . \label{waeff}
\end{equation}
For $n=2$, the left-hand side of (\ref{wtaes}) is equal to
$f_{2}(x)=x-x^{2}$. It is not the case for $n\geq3$. Applying
(\ref{fipoq}) and (\ref{fipoq2}) with $g_{n}(x)$ finally leads to
\begin{equation}
-x\ln{x}\leq
\sum_{s=0}^{n} \wb_{n}^{(s)}x^{s}
\, , \label{wbes}
\end{equation}
where
\begin{equation}
\wb_{n}^{(0)}=1-\sum_{s=1}^{n} \frac{\wta_{n}^{(s)}}{s}
\ , \qquad
\wb_{n}^{(1)}=\wta_{n}^{(1)}-1
\, , \qquad
\wb_{n}^{(s)}=\frac{\wta_{n}^{(s)}}{s}
\qquad
(2\leq{s}\leq{n})
\, . \label{wbeff}
\end{equation}
The result (\ref{wbes}) allows us to estimate the Shannon entropy
from above. For $n=2$, the inequality (\ref{wbes}) coincides with
(\ref{tayln1}). For other values of $n$, coefficients of polynomials
in the formulas (\ref{wtaes}) and (\ref{wbes}) do not correspond to
the Taylor scheme. Hence, the estimates by means of (\ref{wtaes})
and (\ref{wbes}) are not very good near the point $x=1$. For points
sufficiently close to $1$ from the left, we prefer (\ref{tayln0})
and (\ref{tayln1}). In regions peripheral to $x=1$, the results
(\ref{wtaes}) and (\ref{wbes}) provide more accurate estimation.
This distinction is especially clear near the point $x=0$.
Nevertheless, any estimate by polynomials will hardly be very
precise for points close to zero from the right. In effect, the
function of interest is not analytic in this point.

Thus, there is an asymmetry in estimating the function
$x\mapsto-x\ln{x}$ in regions close, respectively, to one of the two
least points of the range $x\in[0,1]$. It is also validated by some
numerical inspection. To make estimates better, one shall move
actual points of approximation to the right, where expected errors
are less. It can be made due to estimating the maximal probability
from above \cite{rastdes}. Let the index of the form (\ref{icsdef}) be
given for some integer $n\geq2$, and let $L$ be the number of
nonzero probabilities. It then holds that \cite{rastdes}
\begin{equation}
\underset{j}{\max}\,p_{j}\leq\Upsilon_{L-1}^{(n)}(I^{(n)})
\, , \label{maxki}
\end{equation}
where $\Upsilon_{L-1}^{(n)}(I^{(n)})$ denotes the maximal real root
of the algebraic equation
\begin{equation}
(1-\Upsilon)^{n}+(L-1)^{n-1}\Upsilon^{n}=(L-1)^{n-1}I^{(n)}
\, . \label{curvey}
\end{equation}
When other parameters are fixed, the quantity
$\Upsilon_{L-1}^{(n)}(I^{(n)})$ is increasing and concave with
respect to $I^{(n)}$ \cite{rastdes}.

For $n\geq5$, we are generally unable to write
$\Upsilon_{L-1}^{(n)}(I^{(n)})$ analytically using radicals.
Instead, it can be calculated numerically with any desired accuracy
\cite{rastdes}. For $n=2,3,4$, one can express the answer in a
closed analytic form. The case $n=2$ is answered by \cite{rastmubs}
\begin{equation}
\Upsilon_{L-1}^{(2)}(I^{(2)})=\frac{1}{L}
\left(1+\sqrt{L-1}\,\sqrt{L\,I^{(2)}-1}\,\right)
 . \label{upst2}
\end{equation}
Due to our findings, the following statement takes place.

\newtheorem{shn1}[shn01]{Proposition}
\begin{shn1}\label{res1}
Let $I^{(s)}(\vcp)$ be given for all $s=2,\ldots,n$. Then the
Shannon entropy satisfies
\begin{equation}
\sum_{s=1}^{n} a_{n}^{(s)}\Upsilon^{1-s}I^{(s)}(\vcp)
-\ln\Upsilon
\leq{H}_{1}(\vcp)\leq
\sum_{s=0}^{n} b_{n}^{(s)}\Upsilon^{1-s}I^{(s)}(\vcp)-
\ln\Upsilon
\, , \label{twotay}
\end{equation}
where $\Upsilon=\Upsilon_{L-1}^{(n)}(I^{(n)})$ and the coefficients
$a_{n}^{(s)}$ and $b_{n}^{(s)}$ are defined by (\ref{aeff}) and
(\ref{boeff}), respectively. It also holds that
\begin{align}
\sum_{s=1}^{n}\wta_{n}^{(s)}\Upsilon^{1-s}I^{(s)}(\vcp)
-\ln\Upsilon\leq{H}_{1}(\vcp)\leq
\sum_{s=0}^{n}\wb_{n}^{(s)}\Upsilon^{1-s}I^{(s)}(\vcp)-
\ln\Upsilon
\, , \label{twocheb}
\end{align}
where the coefficients are defined by (\ref{waeff}) and (\ref{wbeff}).
\end{shn1}

{\bf Proof.}
The points of approximation $x_{j}=p_{j}/\Upsilon$ lie
in the interval $x\in[0,1]$ due to (\ref{maxki}). Substituting
$p_{j}=\Upsilon{x}_{j}$ into (\ref{shdef}) results in the expression
\begin{equation}
H_{1}(\vcp)=\!{}-\ln\Upsilon-\Upsilon\sum_{j=1}^{L}x_{j}\ln{x}_{j}
\, . \label{h1ux}
\end{equation}
Using (\ref{tayln0}) and (\ref{tayln1}) in all the points $x_{j}$,
we obtain after summation that
\begin{equation}
\sum_{s=1}^{n} a_{n}^{(s)}\,\frac{I^{(s)}(\vcp)}{\Upsilon^{s}}
\leq-\sum_{j=1}^{L}x_{j}\ln{x}_{j}\leq
\sum_{s=0}^{n} b_{n}^{(s)}\,\frac{I^{(s)}(\vcp)}{\Upsilon^{s}}
\ . \label{taytay}
\end{equation}
Combining (\ref{h1ux}) with (\ref{taytay}) completes the proof of
(\ref{twotay}). Similarly, the two-sided estimate (\ref{twocheb})
follows from (\ref{wtaes}) and (\ref{wbes}).
$\blacksquare$

The statement of Proposition \ref{res1} provides two-sided estimates
on the Shannon entropy, when several indices of the form
(\ref{icsdef}) are given exactly. In the next section, we will use
(\ref{twotay}) and (\ref{twocheb}) to derive uncertainty and
certainty relations for POVMs assigned to several quantum designs of
degree $3$, $5$, and $7$. They are quantum counterparts of some
spherical designs in three dimensions listed in \cite{hardin96}.
Quantum designs have found a lot of attention due to potential
applications in emerging technologies of information processing.
They also allow us to show a good accuracy of new estimates for a
sufficient number of outcomes. A utility of both the formulas
(\ref{twotay}) and (\ref{twocheb}) will be exemplified as well.
Finally, we use ({\ref{yleqg}) to give an inequality between the
Shannon entropy and Tsallis entropies of integer degree $s\geq2$,
viz.,
\begin{equation}
H_{1}(\vcp)\geq\frac{(-1)^{n}}{2n^{2}}\,\sum_{s=2}^{n} c_{n}^{(s)}H_{s}(\vcp)
\, . \label{tsan1}
\end{equation}
It is important as holding irrespectively to the number of events.

\section{Uncertainty and certainty relations for design-structured POVMs}\label{sec3}

This section is devoted to complementarity relations for POVMs
assigned to a quantum design. When several indices of the form
(\ref{icsdef}) are given, uncertainty and certainty relations
directly follow from the results of the previous section. Let us
recall briefly the required material concerning quantum designs.
Dealing with rays in $d$-dimensional Hilbert space $\hh$, one
selects $K$ unit vectors $|\phi_{k}\rangle$ with the following
property. For all real polynomials $\calp_{t}$ of degree at most $t$
it holds that \cite{scottjpa}
\begin{equation}
\frac{1}{K^{2}}\sum_{j,k=1}^{K}
\calp_{t}\Bigl(\bigl|\langle\phi_{j}|\phi_{k}\rangle\bigr|^{2}\Bigr)=
\int\!\int\xdif\mu(\psi)\,\xdif\mu(\psi^{\prime})\>
\calp_{t}\Bigl(\bigl|\langle\psi|\psi^{\prime}\rangle\bigr|^{2}\Bigr)
\, . \label{tdesdf}
\end{equation}
By $\mu(\psi)$, one denotes the unique unitarily invariant
probability measure on the corresponding complex projective space
induced by the Haar measure. Then unit vectors $|\phi_{k}\rangle$
are said to form a quantum $t$-design. Quantum designs have useful
formal properties posed as follows. Let $\psymt$ be the projector
onto the symmetric subspace of $\hh^{\otimes{t}}$. It holds that
\cite{scottjpa}
\begin{equation}
\frac{1}{K}\,\sum_{k=1}^{K}
|\phi_{k}\rangle\langle\phi_{k}|^{\otimes{t}}=\cald_{d}^{(t)}\psymt
\, , \label{topys}
\end{equation}
where $\cald_{d}^{(t)}$ denotes the inverse of
$\tr\bigl(\psymt\bigr)$, namely,
\begin{equation}
\cald_{d}^{(t)}=\binom{d+t-1}{t}^{\!-1}
=\frac{t!\,(d-1)!}{(d+t-1)!}
\ . \label{indim}
\end{equation}
At the given $t$, we can rewrite (\ref{topys}) for all positive
integers $s\leq{t}$. Substituting $s=1$ leads to the formula
\begin{equation}
\frac{d}{K}\,\sum_{k=1}^{K}
|\phi_{k}\rangle\langle\phi_{k}|=\pen_{d}
\, . \label{comprel}
\end{equation}
Thus, unit vectors $|\phi_{k}\rangle$ allow us to build to a
resolution of the identity in $\hh$. Also, there may be several
resolutions assigned to the given $t$-design. We will call them
design-structured POVMs. The case of single one deals with the
complete set $\cle$ consisting of operators
\begin{equation}
\me_{k}=\frac{d}{K}\>
|\phi_{k}\rangle\langle\phi_{k}|
\, . \label{mkdef}
\end{equation}
Sometimes, the set of $M$ rank-one POVMs $\bigl\{\cle^{(m)}\bigr\}_{m=1}^{M}$
can be assigned to the given quantum design. Each of POVMs
consists of $\ell$ operators of the form
\begin{equation}
\me_{j}^{(m)}=\frac{d}{\ell}\>
|\phi_{j}^{(m)}\rangle\langle\phi_{j}^{(m)}|
\, . \label{mejdf}
\end{equation}
The integers $\ell$ and $M$ are connected by $K=\ell{M}$.

If the state of interest is described by density matrix $\bro$, then
the probability of $j$th outcome is equal to
\begin{equation}
p_{j}(\cle^{(m)};\bro)=\frac{d}{\ell}\,\langle\phi_{j}^{(m)}|\bro|\phi_{j}^{(m)}\rangle
\, . \label{probk}
\end{equation}
Substituting these probabilities into (\ref{shdef}) gives the
entropy $H_{1}(\cle^{(m)};\bro)$. It follows from (\ref{topys}) that
\cite{guhne20}
\begin{equation}
\frac{1}{K}\,\sum_{k=1}^{K}
\langle\phi_{k}|\bro|\phi_{k}\rangle^{s}=\cald_{d}^{(s)}\,\tr\bigl(\bro^{\otimes{s}}\psyms\bigr)
\, . \label{indet}
\end{equation}
Combining (\ref{probk}) with (\ref{indet}) then gives
\begin{equation}
\sum_{m=1}^{M}\sum_{j=1}^{\ell}p_{j}(\cle^{(m)};\bro)^{s}=
\left(\frac{d}{\ell}\right)^{\!s}\,\sum_{k=1}^{K}\langle\phi_{k}|\bro|\phi_{k}\rangle^{s}=
K\ell^{-s}d^{\,s}\,\cald_{d}^{(s)}\,\tr\bigl(\bro^{\otimes{s}}\psyms\bigr)
\, , \label{mindet}
\end{equation}
where $s=2,\ldots,t$. This fact can be used instead of
(\ref{tdesdf}) to verify quantum designs \cite{guhne20}. When a
single POVM is assigned, one has $\ell=K$ and
\begin{equation}
\sum_{k=1}^{K}p_{k}(\cle;\bro)^{s}=
K^{1-s}d^{\,s}\,\cald_{d}^{(s)}\,\tr\bigl(\bro^{\otimes{s}}\psyms\bigr)
\, . \label{indek}
\end{equation}
The authors of \cite{guhne20,cirac18}
resolved the question of how to express
$\tr\bigl(\bro^{\otimes{s}}\psyms\bigr)$ as a sum of monomials of
the moments of $\bro$. On the other hand, such expressions allow us
to get several moments of $\bro$ from quantities of the form
\begin{align}
\bar{\beta}_{\ell}^{(s)}(\bro)&=\ell^{1-s}d^{\,s}\,\cald_{d}^{(s)}\,\tr\bigl(\bro^{\otimes{s}}\psyms\bigr)
\, , \label{betn}\\
\bar{\beta}^{(s)}(\bro)&=K^{1-s}d^{\,s}\,\cald_{d}^{(s)}\,\tr\bigl(\bro^{\otimes{s}}\psyms\bigr)
\, . \label{betk}
\end{align}
For $s=0,1$, we, respectively, have $\bar{\beta}^{(0)}(\bro)=K$ and $\bar{\beta}^{(1)}(\bro)=1$.
The term (\ref{betk}) is obtained from
(\ref{betn}) for $\ell=K$. For a pure state
$\bro=|\psi\rangle\langle\psi|$, the trace in (\ref{betn}) is equal
to $1$ so that
\begin{equation}
\bar{\beta}_{\ell}^{(s)}\bigl(|\psi\rangle\langle\psi|\bigr)=
\ell^{1-s}d^{\,s}\,\cald_{d}^{(s)}
\, . \label{betp}
\end{equation}
The formulas (\ref{mindet}) and (\ref{indek}) are the base to pose
entropic uncertainty relations for design-structured POVMs.
Uncertainty relations in terms of generalized entropies were
considered in \cite{guhne20,rastdes}. Findings of the previous
section allow one to obtain uncertainty and certainty relations in
terms of the Shannon entropy.

\newtheorem{shn2}[shn01]{Proposition}
\begin{shn2}\label{res2}
Let $M$ rank-one POVMs $\cle^{(m)}$, each with $\ell$ elements of
the form (\ref{mejdf}), be assigned to a quantum $t$-design
$\bigl\{|\phi_{k}\rangle\bigr\}_{k=1}^{K}$ in $d$ dimensions. It
then holds that
\begin{align}
\sum_{s=1}^{t} a_{t}^{(s)}\Upsilon^{1-s}\bar{\beta}_{\ell}^{(s)}(\bro)
-\ln\Upsilon
&\leq\frac{1}{M}\sum_{m=1}^{M}H_{1}(\cle^{(m)};\bro)
\leq
\sum_{s=0}^{t} b_{t}^{(s)}\Upsilon^{1-s}\bar{\beta}_{\ell}^{(s)}(\bro)-
\ln\Upsilon
\, , \label{potay}\\
\sum_{s=1}^{t}\wta_{t}^{(s)}\Upsilon^{1-s}\bar{\beta}_{\ell}^{(s)}(\bro)
-\ln\Upsilon
&\leq\frac{1}{M}\sum_{m=1}^{M}H_{1}(\cle^{(m)};\bro)
\leq
\sum_{s=0}^{t}\wb_{t}^{(s)}\Upsilon^{1-s}\bar{\beta}_{\ell}^{(s)}(\bro)
-\ln\Upsilon
\, , \label{poche}
\end{align}
where $\Upsilon=\min\bigl\{M\,\Upsilon_{K-1}^{(t)}\bigl(\bar{\beta}^{(t)}(\bro)\bigr),1\bigr\}$.
\end{shn2}

{\bf Proof.}
For the case of $M$ POVMs, the left-hand side of (\ref{mindet}) is
actually the sum of $s$-indices for all POVMs. For all
$k=1,\ldots,K$, one has
\begin{equation}
\frac{d}{K}\,\langle\phi_{k}|\bro|\phi_{k}\rangle
=p_{k}(\cle;\bro)\leq\Upsilon_{K-1}^{(t)}\bigl(\bar{\beta}^{(t)}(\bro)\bigr)
\, , \nonumber
\end{equation}
whence
\begin{equation}
\frac{d}{\ell}\,\langle\phi_{j}^{(m)}|\bro|\phi_{j}^{(m)}\rangle
=p_{j}(\cle^{(m)};\bro)\leq{M}\,\Upsilon_{K-1}^{(t)}\bigl(\bar{\beta}^{(t)}(\bro)\bigr)
\, \label{pjup}
\end{equation}
due to $K=\ell{M}$. To each of the $M$ entropies
$H_{1}(\cle^{(m)};\bro)$, we apply the two-sided estimate
(\ref{twotay}) with maximal power $t$ and the defined $\Upsilon$ so
that
\begin{equation}
\sum_{s=1}^{t} a_{t}^{(s)}\Upsilon^{1-s}I^{(s)}(\vcp^{(m)})
-\ln\Upsilon
\leq{H}_{1}(\cle^{(m)};\bro)
\leq
\sum_{s=0}^{t} b_{t}^{(s)}\Upsilon^{1-s}I^{(s)}(\vcp^{(m)})-
\ln\Upsilon
\, . \label{potaym}
\end{equation}
It is seen from (\ref{mindet}) that
\begin{equation}
\sum_{m=1}^{M}I^{(s)}(\vcp^{(m)})=
K\ell^{-s}d^{\,s}\,\cald_{d}^{(s)}\,\tr\bigl(\bro^{\otimes{s}}\psyms\bigr)=
M\bar{\beta}_{\ell}^{(s)}(\bro)
\, . \label{sindet}
\end{equation}
Summing (\ref{potaym}) with respect to $m$ and substituting
(\ref{sindet}), we get (\ref{potay}) multiplied by common factor
$M$. Similar reasons allow us to derive (\ref{poche}) from
(\ref{twocheb}).
$\blacksquare$

Thus, we have obtained two families of complementarity relations for
the Shannon entropy averaged over all $M$ POVMs $\cle^{(m)}$. When
single POVM $\cle$ with $K$ elements (\ref{mkdef}) is assigned to
the given $t$-design, the relations (\ref{potay}) and (\ref{poche}),
respectively, reduce to
\begin{align}
\sum_{s=1}^{t} a_{t}^{(s)}\Upsilon^{1-s}\bar{\beta}^{(s)}(\bro)
-\ln\Upsilon
&\leq{H}_{1}(\cle;\bro)\leq
\sum_{s=0}^{t} b_{t}^{(s)}\Upsilon^{1-s}\bar{\beta}^{(s)}(\bro)-
\ln\Upsilon
\, , \label{kpotay}\\
\sum_{s=1}^{t}\wta_{t}^{(s)}\Upsilon^{1-s}\bar{\beta}^{(s)}(\bro)
-\ln\Upsilon
&\leq{H}_{1}(\cle;\bro)
\leq
\sum_{s=0}^{t}\wb_{t}^{(s)}\Upsilon^{1-s}\bar{\beta}^{(s)}(\bro)
-\ln\Upsilon
\, . \label{kpoche}
\end{align}
We shall below consider several examples concerning mainly
(\ref{kpotay}) and (\ref{kpoche}). They are a reflection of
sufficiently strong restrictions imposed on measurement statistics
for design-structured POVMs. Hence, complementarity relations for
various entropic functions follow. Uncertainty relations in terms of
generalized entropies were considered in \cite{guhne20,rastdes}. In
a sense, the above discussion complements the analysis by adding
relations in terms of the Shannon entropy. For a pure state, the
left-hand sides of (\ref{potay}) and (\ref{poche}) lead to
\begin{align}
\frac{1}{M}\sum_{m=1}^{M} H_{1}\bigl(\cle^{(m)};|\psi\rangle\langle\psi|\bigr)
&\geq\sum_{s=1}^{t} a_{t}^{(s)}(\Upsilon\ell)^{1-s}d^{\,s}\,\cald_{d}^{(s)}
-\ln\Upsilon
\, , \label{popit}\\
\frac{1}{M}\sum_{m=1}^{M} H_{1}\bigl(\cle^{(m)};|\psi\rangle\langle\psi|\bigr)
&\geq\sum_{s=1}^{t}\wta_{t}^{(s)}(\Upsilon\ell)^{1-s}d^{\,s}\,\cald_{d}^{(s)}
-\ln\Upsilon
\, , \label{popich}
\end{align}
where
$\Upsilon=\min\bigl\{M\,\Upsilon_{K-1}^{(t)}\bigl(K^{1-t}d^{\,t}\,\cald_{d}^{(t)}\bigr),1\bigr\}$.

The formulas (\ref{popit}) and (\ref{popich}) are interesting due to
the following. In many applications of uncertainty relations, we ask
for restrictions that hold for all states. Such state-independent
relations are often reached with substituting a pure state. In
typical cases, we deal with several mutually unbiased bases or POVMs assigned to a
quantum design. Here, a total set of used vectors is overcomplete
so that generated probabilities are distributed more uniformly for
sufficiently mixed states. Of course, entropic functions then take
larger values on average. This behavior coincides with increasing
$\Upsilon_{K-1}^{(t)}(\beta)$ with respect to $\beta$
\cite{rastdes}. Quantities of the form (\ref{betn}) and (\ref{betk})
are clearly maximal for pure states. Combining this with
(\ref{pjup}) shows that a diapason of probability values shortens
with a growth of state mixedness. So, the choice of a pure state
leads to lower entropic bounds that hold for all states. The fact
was too shown for R\'{e}nyi-entropy uncertainty relations
\cite{rastdes}. Due to the applied methods of derivation, the
validity of (\ref{popit}) and (\ref{popich}) for all states is not
easy to gain analytically. Nevertheless, we will see this physically
natural behavior with all the examples. By the way, the last term in
the right-hand sides of both (\ref{popit}) and (\ref{popich})
exhibits such a behavior. In any case, the desired fact can be
checked by inspection in each concrete example of design-structured
POVMs. At the end of this section, we will apply
(\ref{popit}) and (\ref{popich}) as state-independent formulations.
The state-independent upper bound is merely $\ln\ell$ for each of
$M$ POVMs and $\ln{K}$ for a single POVM.

\begin{figure*}
\centering \includegraphics[height=7.7cm]{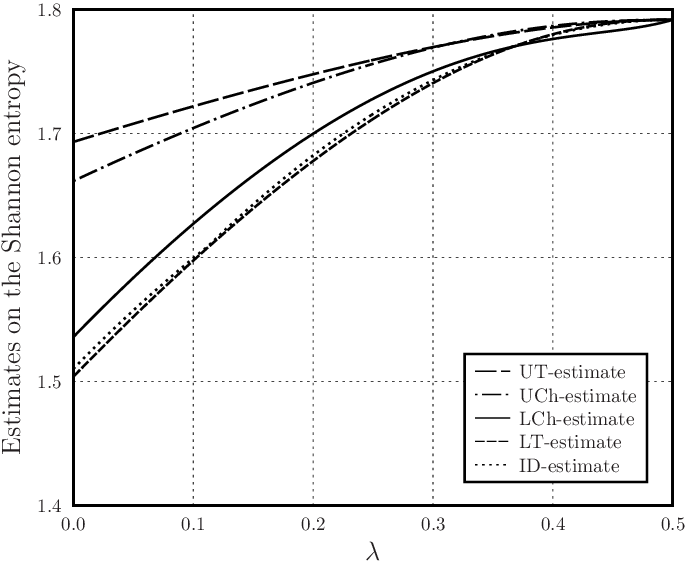}
\caption{\label{fig2} Estimates on the Shannon entropy versus $\lambda$ for the $3$-design with six vertices.}
\end{figure*}

Let us compare the above finding with relations previously published
in the literature. Here, we quote some results of the papers
\cite{harr2001,molm09}. One of the results of Harremo\"{e}s and
Tops{\o}e \cite{harr2001} can be reformulated as follows
\cite{molm09}: For probability distribution $\vcp$ with $K$
probabilities, it holds that
\begin{equation}
H_{1}(\vcp)\geq\ln(k+1)+k\ln\biggl(\frac{k+1}{k}\biggr)-k(k+1)\ln\biggl(\frac{k+1}{k}\biggr)\,I^{(2)}(\vcp)
\, , \label{tomolm}
\end{equation}
where $k$ is integer and $1\leq{k}\leq{K}-1$. To increase the lower
bound, we take $k$ depending on the actual value of $I^{(2)}(\vcp)$.
The choice $k=K-1$ is optimal only for states sufficiently close to
the maximally mixed one. In the following examples, the right-hand
side of (\ref{tomolm}) is maximized with respect to integer $k$.

Let us proceed to examples of applications to concrete quantum
designs in two dimensions. A short description of these designs in
terms of components of the Bloch vector is given in \cite{guhne20}.
This vector comes to one of the vertices forming some polyhedron. To
characterize the qubit density matrix, its minimal eigenvalue
$\lambda$ is used. It is instructive to visualize the two-sided
estimates (\ref{potay}) and (\ref{poche}) together with
(\ref{tomolm}). To avoid bulky legends on figures, the following
notation will be utilized. By ``LT-estimate'' and ``UT-estimate,''
we mean the left- and right-hand sides of (\ref{potay}),
respectively. They are based on approximation by polynomials, whose
coefficients are due to the Taylor scheme. The terms
``LCh-estimate'' and ``UCh-estimate,'' respectively, refer to the
left- and right-hand sides of (\ref{poche}). Due to (\ref{yleqg}),
such estimates use polynomials with coefficients linked to
coefficients of the shifted Chebyshev polynomials. The 
right-hand side of (\ref{tomolm}) following from information
diagrams will be referred to as ``ID-estimate.'' We will mainly
focus on the case of single assigned POVMs.

\begin{figure*}
\centering \includegraphics[height=7.7cm]{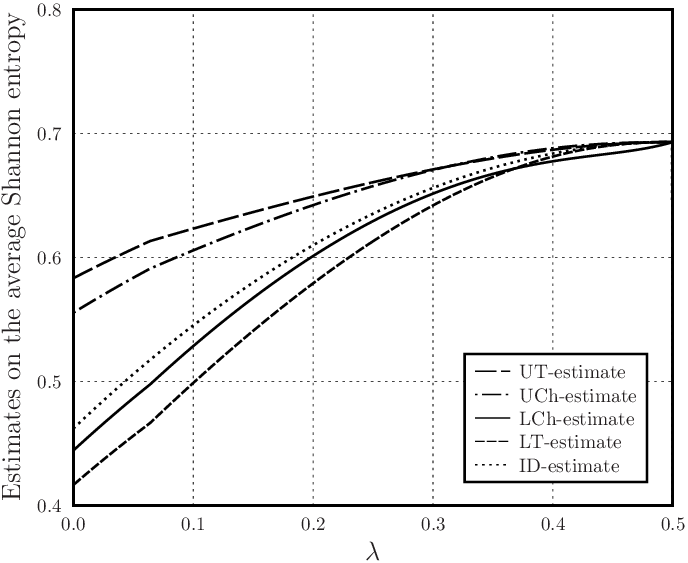}
\caption{\label{fig3} Estimates on the average Shannon entropy versus $\lambda$ for the three mutually unbiased bases in two dimensions.}
\end{figure*}

Let us begin with the $3$-design with $K=6$ vertices forming an
octahedron. In Figure \ref{fig2}, we plot the three lower estimates
and the two upper ones as functions of $\lambda$. Of course, the
corresponding Shannon entropy varies in sufficiently restricted
diapason. The two-sided estimate (\ref{kpotay}) is better only for
states very close to the maximally mixed one. We also notice that
ID-estimate almost coincides with the left-hand side of
(\ref{kpotay}) but is better for almost all $\lambda$. Thus, the result
(\ref{tomolm}) well converts the actual value of (\ref{ic2def}) into
estimating the Shannon entropy from below. In fact, the proposed
estimates are not very good for degree $3$. Nevertheless, the
formula (\ref{kpoche}) shows considerably better results for pure
states and states with low mixedness. For pure states, the
difference between LCh-estimate and ID-estimate is more then one
fifth of the difference between UCh-estimate and LCh-estimate. The
latter describes a diapason in which the Shannon entropy varies.
Overall, LCh- and UCh-estimates are better in a sufficiently wide range of
$\lambda$. For the maximally mixed state, the five curves are all
converging at one point.

In the example with the octahedron, the $3$-design of interest is formed
by eigenstates of the Pauli matrices. Here, one deals with the
complete set of three mutually unbiased bases in two dimensions.
Entropic uncertainty relations for such bases were extensively
studied. Using (\ref{tomolm}), the authors of \cite{molm09} studied
entropic uncertainty relations for a set of mutually unbiased bases.
For $M=3$ and $d=2$, their main result reads as
\begin{equation}
\frac{1}{3}\sum_{m=1}^{3}H_{1}(\cle^{(m)};\bro)\geq
\frac{2-\tr(\bro^{2})}{3}\,\ln4
 . \label{molm}
\end{equation}
In this example, the right-hand side of (\ref{molm}) will be
referred to as ID-estimate. The five estimates on the average
Shannon entropy are presented in Figure \ref{fig3}. In this example,
ID-estimate provides a better lower bound. For pure states, the
difference between ID-estimate and LCh-estimate is almost one fifth
of the difference between UCh-estimate and ID-estimate. It was
already noticed that our estimates seem to be insufficient for
degree $3$. Another origin of comparatively poor results of
(\ref{poche}) is due to the fact that
$3\,\Upsilon_{5}^{(3)}\bigl(\bar{\beta}^{(3)}(\bro)\bigr)>1$ for
states with sufficiently low mixedness. A more accurate way to
estimate the average maximal probability from above is based on the
inequality \cite{rastdes}
\begin{equation}
\frac{1}{M}\sum_{m=1}^{M}\underset{j}{\max}\,p_{j}(\cle^{(m)};\bro)
\leq\Upsilon_{\ell-1}^{(t)}\bigl(\bar{\beta}_{\ell}^{(t)}(\bro)\bigr)
\, . \label{apmax}
\end{equation}
The latter should be used here with $t=3$ and $\ell=2$.
Unfortunately, methods of deriving (\ref{potay}) and (\ref{poche})
are such that one is hardly able to use just (\ref{apmax}). The
right-hand side of (\ref{tomolm}) has a simple analytical structure
such that results of the form (\ref{molm}) follow. This example
shows that the proposed two-sided estimates provide better results
in application to a single entropy. As was already mentioned, only
the sum of $s$-indices up to $s=t$ is given by (\ref{mindet}).

\begin{figure*}
\centering \includegraphics[height=7.7cm]{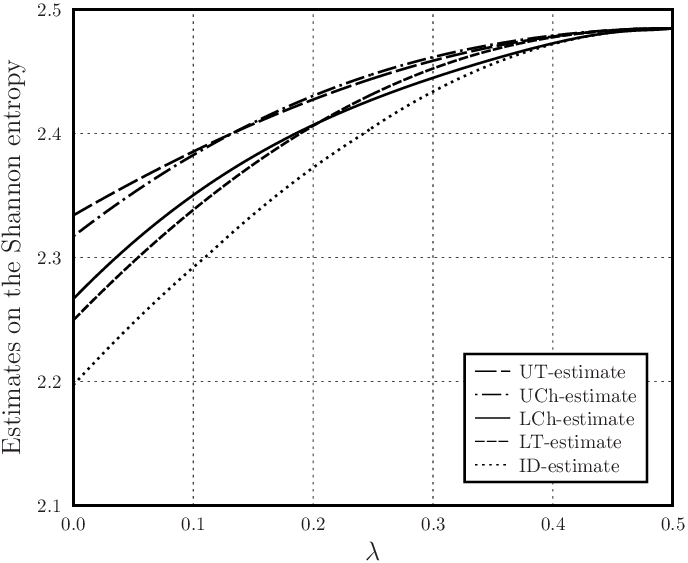}
\caption{\label{fig4} Estimates on the Shannon entropy versus $\lambda$ for the $5$-design with $12$ vertices.}
\end{figure*}

The following example is the $5$-design with $K=12$ vertices forming
an icosahedron. The three lower bounds and two upper bounds versus
$\lambda$ are shown in Figure \ref{fig4}. Of course, allowed changes
of the corresponding Shannon entropy are enough limited in their
value. Close to the maximally mixed state, the two-sided estimate
(\ref{kpotay}) is better than (\ref{kpoche}). Moreover, we see a wider
region in which the result (\ref{kpotay}) leads to stronger bounds.
For states with low mixedness, the left-hand sides of (\ref{kpotay})
and (\ref{kpoche}) are both considerably stronger than the lower
estimate obtained from the information diagrams. For pure states,
the difference between LCh-estimate and ID-estimate is almost $1.4$
times the difference between UCh-estimate and LCh-estimate. In other
words, ID-estimate here is far from actual entropic values. It is
insufficient even in comparison with (\ref{kpotay}). We could expect
this, since ID-estimate uses only the index of coincidence
(\ref{ic2def}), whereas new estimates take into account indices of
the form (\ref{icsdef}) for $s=2,3,4,5$. Similarly to the previous
examples, all the five curves converge at one point for the
maximally mixed state.

Let us consider also the $5$-design with $K=30$ vertices forming an
icosidodecahedron. On the average, allowed values of the Shannon
entropy are larger in view of an increased number of outcomes. The five
curves shown in Figure \ref{fig5} curves form a picture quite
similar to the previous example. The interval $\lambda\in[0,1/2]$
is divided into two approximately equal parts. The former is where
the two-sided estimate (\ref{kpoche}) gives better results, and the
latter is a domain for the use of (\ref{kpotay}). As before, for
pure states ID-estimate is sufficiently far from optimality. In
effect, the difference between LCh-estimate and ID-estimate is
comparable with the difference between UCh-estimate and
LCh-estimate. Nevertheless, this distinction is less than in the
previous case with $12$ vertices. Close to the maximally mixed
state, all of the five curves become coinciding.

\begin{figure*}
\centering \includegraphics[height=7.7cm]{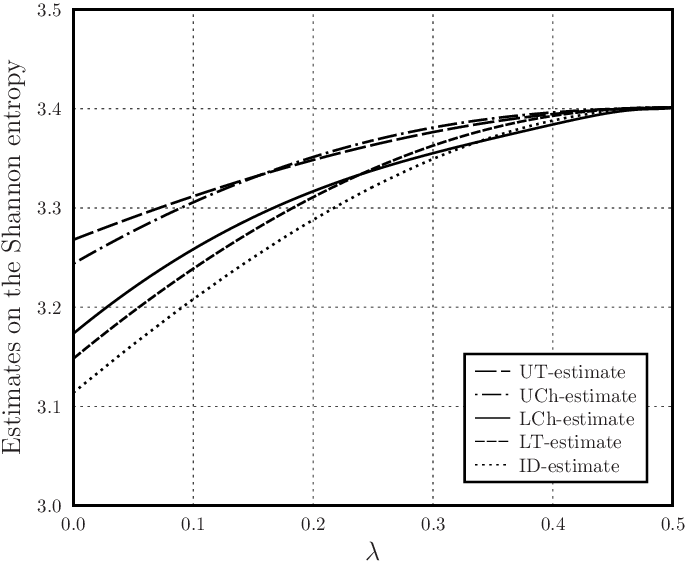}
\caption{\label{fig5} Estimates on the Shannon entropy versus $\lambda$ for the $5$-design with $30$ vertices.}
\end{figure*}

Finally, we test the discussed estimates with the $7$-design
obtained from a deformed snub cube. The regular snub cube has $60$
equal edges, and its $24$ vertices form a $3$-design. Moving these
vertices slightly, one can obtain a 7-design \cite{hardin96}. The
resulting $7$-design was first found by McLaren \cite{laren63}. In
Figure \ref{fig6}, we plot the estimates of interest for the McLaren
design. A range for possible values of the Shannon entropy is more
narrow than in the previous examples. Another consequence of
increasing $t$ is that ID-estimate becomes insufficient for states
beyond a vicinity of the maximally mixed one. For pure states, the
difference between LCh-estimate and ID-estimate is almost three
times the difference between UCh-estimate and LCh-estimate. All five
curves tend to coincide on the right. The two-sided estimate
(\ref{kpoche}) gives better results for states with low mixedness.
Otherwise, the two-sided estimate (\ref{kpotay}) is preferable.
Distinctions between these estimates are less than in the previous
examples.

Design-structured POVMs can be used for estimating the von Neumann
entropy of a quantum state, viz.,
\begin{equation}
\rmh_{1}(\bro):=\!{}-\tr(\bro\ln\bro)
\, . \label{vnbm}
\end{equation}
Let us restrict a consideration to a single assigned POVM that gives
quantities of the form (\ref{betk}) for $s=2,\ldots,t$. Then the
trace $\tr(\bro^{2})$ is gained from $\bar{\beta}^{(2)}(\bro)$, the
trace $\tr(\bro^{3})$ is gained from $\bar{\beta}^{(3)}(\bro)$ and
the known $\tr(\bro^{2})$, and so on. In this way, we find such
traces with integer powers of $\bro$ up to $\tr(\bro^{\!\;t})$. We
can interpret (\ref{vnbm}) as the Shannon entropy calculated with
eigenvalues of $\bro$. It then follows from (\ref{twotay}) and
(\ref{twocheb}) that
\begin{align}
\sum_{s=1}^{t} a_{t}^{(s)}\Lambda^{1-s}\,\tr(\bro^{s})
-\ln\Lambda
&\leq\rmh_{1}(\bro)\leq{b}_{t}^{(0)}\Lambda{d}+
\sum_{s=1}^{t} b_{t}^{(s)}\Lambda^{1-s}\,\tr(\bro^{s})-
\ln\Lambda
\, , \label{potayn}\\
\sum_{s=1}^{t}\wta_{t}^{(s)}\Lambda^{1-s}\,\tr(\bro^{s})
-\ln\Lambda
&\leq\rmh_{1}(\bro)
\leq\wb_{t}^{(0)}\Lambda{d}+
\sum_{s=1}^{t}\wb_{t}^{(s)}\Lambda^{1-s}\,\tr(\bro^{s})
-\ln\Lambda
\, , \label{pochen}
\end{align}
where $\Lambda=\Upsilon_{d-1}^{(t)}\bigl(\tr(\bro^{\!\;t})\bigr)$.
Here, the quantity $\Lambda$ estimates the eigenvalues of $\bro$
from above. When $t\geq{d}$, measurement statistics with many trials
lead to determining eigenvalues of $\bro$. Otherwise, only ranges of
available values could be mentioned. Yet, the formulas
(\ref{potayn}) and (\ref{pochen}) allow us to estimate promptly the
von Neumann entropy of input state from both the sides. For example,
Figure \ref{fig7} shows the von Neumann entropy and its estimates
for the qubit case with $t=5$. The $5$-design can be formed by
icosahedron vertices marked on the Bloch sphere. Except for states
with low mixedness, an accuracy of estimation is very good.

\begin{figure*}
\centering \includegraphics[height=7.7cm]{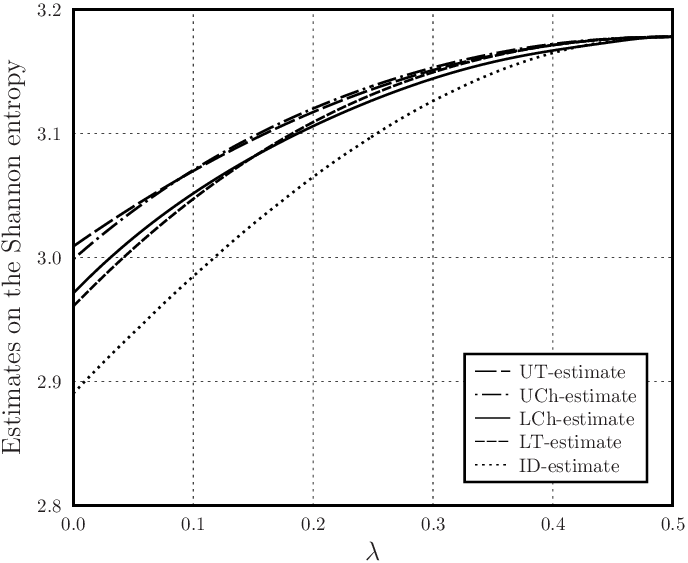}
\caption{\label{fig6} Estimates on the Shannon entropy versus $\lambda$ for the $7$-design with $24$ vertices.}
\end{figure*}

The above uncertainty relations for design-structured POVMs can be
used to formulate quantum steering inequalities. This question was
already addressed in \cite{guhne20,rastdes}. So, we will discuss
steering inequalities very shortly. For more details, see the review
\cite{ucno20} and references therein. Alice and Bob share a
bipartite quantum state $\bro_{AB}$ and repeat this any number of
times \cite{wisem07}. Alice performs on her subsystem a measurement
chosen from the set of POVMs $\bigl\{\clf^{(m)}\bigr\}_{m=1}^{M}$.
Hence, the actual state of Bob's subsystem is conditioned on Alice's
result. This conditioned state is subjected to a measurement chosen
accordingly from the set $\bigl\{\cle^{(m)}\bigr\}_{m=1}^{M}$. The
conditional entropies $H_{1}\bigl(\cle^{(m)}|\clf^{(m)}\bigr)$ are
calculated due to (\ref{csea}) with the use of generated
probabilities and classical side information from Alice. The authors
of \cite{brun18} examined the question of how to derive quantum
steering inequalities on the base of entropic uncertainty relations.
One of the conditions implies that state-independent uncertainty
relations should be used. Other conditions are clearly valid for the
standard entropic functions. As was already mentioned, validity of
the lower entropic bounds (\ref{popit}) and (\ref{popich}) for all
states should be checked in each concrete example. If they actually
provide a state-independent formulation, then we have the steering
inequalities
\begin{align}
\frac{1}{M}\sum_{m=1}^{M} H_{1}\bigl(\cle^{(m)}|\clf^{(m)}\bigr)
&\geq\sum_{s=1}^{t} a_{t}^{(s)}(\Upsilon\ell)^{1-s}d^{\,s}\,\cald_{d}^{(s)}
-\ln\Upsilon
\, , \label{posit}\\
\frac{1}{M}\sum_{m=1}^{M} H_{1}\bigl(\cle^{(m)}|\clf^{(m)}\bigr)
&\geq\sum_{s=1}^{t}\wta_{t}^{(s)}(\Upsilon\ell)^{1-s}d^{\,s}\,\cald_{d}^{(s)}
-\ln\Upsilon
\, , \label{posich}
\end{align}
where
$\Upsilon=\min\bigl\{M\,\Upsilon_{K-1}^{(t)}\bigl(K^{1-t}d^{\,t}\,\cald_{d}^{(t)}\bigr),1\bigr\}$.
Violation of any steering inequalities means steerability of the
tried state. It was previously discussed that the two-sided
estimates (\ref{potay}) and (\ref{poche}) give better results in the
case of single assigned POVM. Similarly, the same feature characterizes
the steering inequalities (\ref{posit}) and (\ref{posich}).

\begin{figure*}
\centering \includegraphics[height=7.7cm]{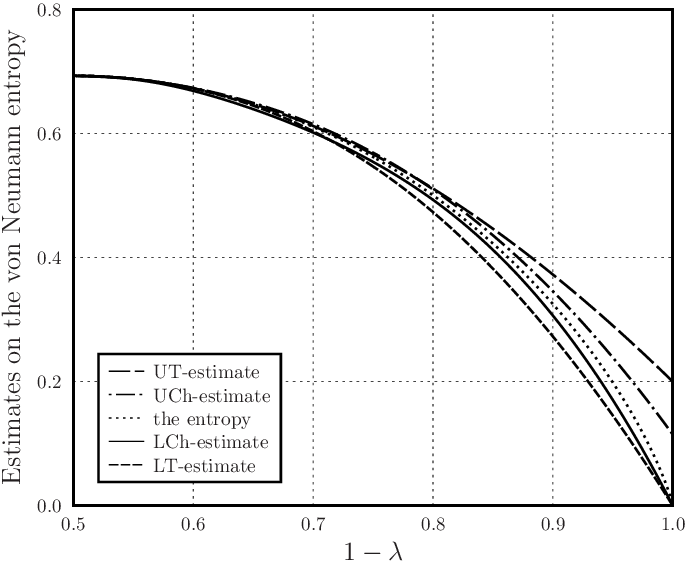}
\caption{\label{fig7} Estimates on the von Neumann entropy of a qubit with $t=5$.}
\end{figure*}

\section{Concluding remarks}\label{sec4}

We considered methods to derive two-sided estimates on the Shannon
entropy by means of polynomial functions. The first way is based on
truncated expansions of the Taylor type. It leads to results that
are intuitively clear and easy to prove. The second way uses
expansions with flexible coefficients. The significance of such
expansions in applied analysis was emphasized by Lanczos
\cite{lanczos}. As a result, we have arrived at a family of
polynomials whose coefficients are connected with the shifted
Chebyshev polynomials. Overall, this way provides more balanced
estimates. On the other hand, the validity of obtained estimates is
not obvious. It turned out that polynomials of moderate degree are
often sufficient to reach enough good results. Any advances to
reveal the nature of polynomial estimates with flexible coefficients
will be welcome.

Recently, quantum designs are actively studied due to constraints of
the form (\ref{mindet}), and their corollaries are useful in quantum
information processing. The derived two-sided estimates were applied
to formulate uncertainty and certainty relations for
design-structured POVMs. A quality of new complementarity relations
is characterized by comparing with the lower entropic bounds
following from information diagrams. In general, new relations lead
to stronger inequalities in application to the case of a single
assigned POVM. Also, an actual degree of the corresponding
polynomial should not be very low. It is natural due to the increased
number of restrictions imposed on generated probabilities. So, the
proposed approach allows us to take into account several given
indices of the form (\ref{icsdef}). It was also exemplified that
further improvements seem to be achievable. Maybe information
diagrams with more indices could be used for these purposes.
However, such diagrams will be complicated to examine.

\appendix

\section{Some facts about Chebyshev polynomials}\label{cheb}

The Chebyshev polynomials of the first kind are defined in terms of
the ordinary generating function (see, e.g., item 22.9.9 in Table
22.9 of \cite{stegun72})
\begin{equation}
\sum\nolimits_{n=0}^{\infty} T_{n}(\xi)\,\tau^{n}=\frac{1-\tau\xi}{1-2\tau\xi+\tau^{2}}
\ . \label{ogenf}
\end{equation}
In the main text, we refer to the shifted Chebyshev polynomials
defined as
\begin{equation}
T_{n}^{*}(x)=T_{n}(2x-1)
\, , \label{shifch}
\end{equation}
where $x\in[0,1]$. The representation reads as (see, e.g., section VII.8
in \cite{lanczos})
\begin{equation}
T_{n}^{*}(x)=\sum\nolimits_{s=0}^{n} c_{n}^{(s)}x^{s}
\, , \qquad
c_{n}^{(s)}=(-1)^{n+s}\,2^{2s-1}
\!\left[
2\binom{n+s}{n-s}-\binom{n+s-1}{n-s}
\right]
 . \label{shchs}
\end{equation}
In particular, we write the two coefficients
\begin{equation}
c_{n}^{(0)}=(-1)^{n}
\, , \qquad
c_{n}^{(1)}=(-1)^{n+1}2n^{2}
\, . \label{cn01}
\end{equation}
Exact values of first coefficients can be found, e.g., in Table
VII of the appendix of \cite{lanczos}.

\section{On the first derivative of (\ref{gansun})}\label{gnex}

Differentiating (\ref{gansun}) with respect to $x$ and substituting
$x=1$, one has
\begin{equation}
g_{n}^{\prime}(1)=\frac{(-1)^{n}}{2n^{2}}\sum_{s=2}^{n} c_{n}^{(s)}=
\frac{(-1)^{n}}{2n^{2}}\,\bigl[\,T_{n}^{*}(1)-c_{n}^{(0)}-c_{n}^{(1)}\bigr]
=\frac{(-1)^{n}-1+2n^{2}}{2n^{2}}
\ , \label{gansun1}
\end{equation}
where we used $T_{n}^{*}(1)=T_{n}(1)=1$. The first derivative at the
point $x=0$ satisfies
\begin{equation}
(-1)^{n+1}2n^{2}g_{n}^{\prime}(0)=
\sum_{s=2}^{n} \,\frac{c_{n}^{(s)}}{s-1}
=\int_{0}^{1}
\frac{T_{n}^{*}(x)-c_{n}^{(0)}-c_{n}^{(1)}x}{x^{2}}\>\xdif{x}
\, . \label{gansun0}
\end{equation}
Multiplying (\ref{gansun0}) by $\tau^{n}$ and summing with respect
to $n$, we have
\begin{equation}
\int_{0}^{1}\left[
\frac{1-\tau\xi}{1-2\tau\xi+\tau^{2}}-
\sum\nolimits_{n=0}^{\infty}\bigl(c_{n}^{(0)}+c_{n}^{(1)}x\bigr)\tau^{n}
\right]\frac{\xdif{x}}{x^{2}}
\, , \label{inint1}
\end{equation}
where $\xi=2x-1$. Substituting (\ref{cn01}), usual calculus allows us to
rewrite (\ref{inint1}) as
\begin{equation}
\sum_{n=0}^{\infty}\tau^{n}\sum_{s=2}^{n} \,\frac{c_{n}^{(s)}}{s-1}
=\frac{4(\tau-\tau^{2})}{(1+\tau)^{3}}\>\ln\frac{1+\tau}{1-\tau}
\ . \label{inintres}
\end{equation}
The expansion of (\ref{inintres}) into power series starts with the
term $\propto\tau^{2}$ and results in
\begin{equation}
\sum_{n=0}^{\infty}\tau^{n}\sum_{s=2}^{n} \,\frac{c_{n}^{(s)}}{s-1}=
(-8)\sum_{q=1}^{\infty}q^{2}(-\tau)^{q} \sum_{\substack{m=1 \\ m\>{\mathrm{odd}}}}^{\infty}\frac{\tau^{m}}{m}
\ . \label{twoser}
\end{equation}
We further find the coefficient of $\tau^{n}$ in (\ref{twoser}).
Combining this with (\ref{gansun0}) leads to (\ref{gp0odd}) and
(\ref{gp0even}).

\end{document}